\documentstyle[12pt,epsfig]{article}

\newcommand{\lsim}{\mathrel{\lower4pt\hbox{$\sim$}}
\hskip-12.5pt\raise1.6pt\hbox{$<$}\;}

\newcommand{\gsim}{\mathrel{\lower4pt\hbox{$\sim$}}
\hskip-12.5pt\raise1.6pt\hbox{$>$}\;}

\newcommand{\nc}{\newcommand}

\nc{\lsp}{\;\;\;\;\;\;\;\;}
\nc{\beq}{\begin{equation}}   \nc{\eeq}{\end{equation}}
\nc{\bea}{\begin{eqnarray}}   \nc{\eea}{\end{eqnarray}}
\nc{\baa}{\begin{array}}      \nc{\eaa}{\end{array}}

\begin{document}
\baselineskip 18pt plus 2pt

\noindent \hspace*{11cm}UCRHEP-T206\\
\noindent \hspace*{11cm}October 1997\\

\begin{center}
{\large\bf Probing New Higgs-Top Interactions at the 
Tree-Level in a Future $e^+e^-$ Collider}~\footnote{Expanded version of a talk presented at 
the International Europhysics Conference on High Energy Physics, EPS-HEP97, August 19-26, 1997, Jerusalem, Israel; Based on works done in collaboration with D. Atwood, G. Eilam, A. Soni and J. Wudka.}
\vspace{1cm}

Shaouly Bar-Shalom
\vspace{1cm}

Department of Physics, University of California, \\ Riverside CA 92521, USA.
\end{center}

\vspace{1cm}

\begin{center}
{\bf Abstract} 
\end{center}
\begin{quote}
The possibility of observing large signatures of new CP-violating and flavor-changing Higgs-Top couplings in a future $e^+e^-$ collider experiments such as $e^+e^- \to t \bar t h,~t \bar t Z$ and $e^+e^- \to t \bar c \nu_e {\bar\nu}_e,~t \bar c e^+e^-$ is discussed. Such, beyond the Standard Model, couplings can occur already at the tree-level within a class of Two Higgs Doublets Models. Therefore, an extremely interesting feature of those reactions is that the CP-violating and flavor-changing effects are governed by tree-level dynamics. These reactions may therefore serve as unique avenues for searching for new phenomena associated with Two Higgs Doublets Models and, as is shown here, could yield statistically significant signals of new physics. We find that the CP-asymmetries 
in $e^+e^- \to t \bar t h,~t \bar t Z$ can reach tens of 
percents, and the flavor changing cross-section 
of $e^+e^- \to t \bar c \nu_e {\bar\nu}_e$ is typically a few fb's,  
for light Higgs mass around the electroweak scale. 
\end{quote}

\newpage

\noindent {2. \bf Introductory Remarks And Two Higgs 
Doublets Models} \\

A future high energy $e^+e^-$ collider running at c.m.\ energies of 0.5--2
TeV,  often referred to as the Next Linear Collider (NLC), is designed, 
in part, to study in detail the nature of the scalar potential \cite{nlc}.
It may be also very useful for a close examination of 
the top quark Yukawa couplings to scalar particle(s) 
which, in turn, may give us a clue about   
the properties of the
scalar particle(s). 

In a NLC with a yearly integrated luminosity of the order of 
$100$ fb$^{-1}$, 
a detailed study of cross-sections 
at the level of $\sim$few fb's may become feasible.    
In particular, 
for top-Higgs systems,
the NLC will enable a detailed examination of new phenomena, 
beyond the Standard Model (SM), associated 
with new CP-violating and flavor-changing (FC) 
top Yukawa couplings to scalar particle(s), 
such as the ones discussed here. 
Indeed, the top quark, being so heavy, $m_t \sim 175$ GeV,    
is the most sensitive to these new interactions.

In the SM, the scalar potential is economically composed of only one
scalar doublet. Even a mild extension of the SM  with an additional scalar
doublet \cite{gunion}, can give  rise to rich new phenomena beyond 
the SM associated with top-Higgs systems, e.g., tree-level CP-violation
\cite{prd53p1162,hepph9707284,atwood} and tree-level 
flavor-changing-scalar (FCS) transitions
\cite{prl79p1217,hou,fc2hdm}, 
in interaction of  neutral scalars with the top quark. 

Here we present three distinct reactions which are very powerful probes 
of the $tth$ and $t \bar ch$ Yukawa couplings. The first two reactions, 
$e^+e^- \to t \bar t h$ and $e^+ e^- \to t \bar t Z$ 
\cite{prd53p1162,hepph9707284}, exhibit large 
CP-violating asymmetries, at the order of tens of percent, already 
at {\it tree-level}. The third is {\it tree-level} 
$t \bar c$ production 
through the $W^+W^-$ fusion process, 
$e^+e^-\to W^+ W^- \nu_e\bar\nu_e \to t \bar c\nu_e\bar\nu_e$
\cite{prl79p1217}, which 
appears to be extremely  
sensitive to a $t \bar ch$ FCS
interaction.     

In the presence of two Higgs doublets the most general Yukawa 
lagrangian can be written as:
\begin{equation}
{\cal L}_Y = U^1_{ij} {\bar q}_{i,L} {\tilde {\phi}}_1 u_{j,R} + 
D^1_{ij} {\bar q}_{i,L} {\phi}_1 d_{j,R} + 
U^2_{ij} {\bar q}_{i,L} {\tilde {\phi}}_2 u_{j,R} + 
D^2_{ij} {\bar q}_{i,L} {\phi}_2 d_{j,R} + {\rm h.c.} \label{yukawa}~,
\end{equation}
\noindent where $\phi_i$ for $i=1,2$ are the two scalar doublets and $U^k_{ij},D^k_{ij}$,
for $k=1,2$, are the Yukawa  couplings matrices which are in general non-diagonal.
Depending  on the assumptions made, one can then obtain different versions
of a Two Higgs Doublet Model (2HDM).  In particular, if one imposes the
 discrete symmetries $\phi_1;\phi_2 \to -\phi_1;\phi_2$  and $d_{i,R};u_{i,R}
 \to -d_{i,R};-u_{i,R}~or~-d_{i,R};u_{i,R}$  one arrives at the so called
 Model~I or Model~II, respectively,  depending on whether the -1/3 and 2/3
 charged quarks are coupled  to the same or to different scalar doublets.
 If, in addition,  these discrete symmetries are softly violated by a  mass-dimension-two
 term in the Higgs potential,  then the real  and imaginary parts of the
 Higgs doublets mix, giving rise to  CP-violating scalar-pseudoscalar mixed
 couplings of a neutral  Higgs to fermions already at the tree-level
 \cite{froggat}. 
On the other hand, if one does not impose the above discrete
 symmetries, one arrives at a most general version of the 2HDM, often called
 Model~III, in which both FCS  transitions and CP-nonconserving interactions
 between the neutral  Higgs particles and fermions are present at tree-level
 (for a recent short review see e.g., \cite{sonirev}).   
 The scalar spectrum of any of the above 2HDM's consists of three neutral 
Higgs 
 and two charged Higgs particles which are not relevant for the 
present discussion. For reasons discussed in the following sections, 
for both the CP-violating effects 
in $e^+e^- \to t \bar t h,t \bar t Z$ 
and the FC effects in 
$e^+e^- \to t \bar c\nu_e\bar\nu_e$, 
only two out of the three neutral Higgs 
of the 2HDM's (i.e., Models II and III) are relevant. 
We denote these two neutral Higgs particles by $h$ and $H$ 
corresponding to the lighter and heavier Higgs-boson, 
respectively. In some instances we denote a neutral Higgs 
by ${\cal H}$, then ${\cal H}=h$ or $H$ is to be understood. 

The ${\cal H} t \bar t$ interaction lagrangian piece of 
a general 2HDM can be written as:  
\begin{equation}
{\cal L}_{{\cal H} tt}= -\frac{g_W}{\sqrt 2} \frac{m_t}{m_W} {\cal H} 
\bar t \left( a_t^{\cal H} + i b_t^{\cal H} \gamma_5 \right) t 
\label{tthcoupling}~,
\end{equation}
\noindent where in Model II used here:
\begin{equation}
a_t^h;a_t^H=\frac{R_{11}}{\sin\beta};\frac{R_{12}}{\sin\beta}~~,~~
b_t^h;b_t^H=\frac{R_{31}}{\tan\beta};\frac{R_{32}}{\tan\beta}
 \label{mod2coup} ~.
\end{equation}
\noindent $\tan\beta \equiv v_u / v_d$ 
and $v_u$($v_d$) is the vacuum-expectation-value (VEV) 
responsible for giving mass to  the up(down) quark. 
$R$ is the neutral Higgs
mixing  matrix which can be parameterized by three Euler angles  
$\alpha_{1,2,3}$ \cite{froggat}. 
Note that in the SM the only couplings in 
Eqs.~\ref{tthcoupling} and \ref{mod2coup},
  of the one neutral Higgs present, are $a_t^h=1/\sqrt {2},b_t^h=0$ and
there is no phase in the $ht \bar t$ interaction lagrangian.  

In Model III with the Cheng-Sher Ansatz (CSA) \cite{csa},  
the couplings of the neutral scalars to fermions are given by
$\xi_{ij}^{U,D}=g_W \left({\sqrt {m_im_j}}/m_W \right) \lambda_{ij}$
and the $t \bar ch$ interaction Lagrangian can then be written as:
\begin{eqnarray}
{\cal L}_{{\cal H}tc} = -\frac{g_W}{\sqrt 2} 
\frac{{\sqrt {m_t m_c}}}{m_W} f^{\cal H} {\cal H} 
\bar t (\lambda_R+i\lambda_I \gamma_5) c  \label{htc}~,
\end{eqnarray}
\noindent where for simplicity we choose $\lambda_{tc}=\lambda_{ct}=\lambda$.\footnote{Existing experimental information does not provide
any useful constraints on $\lambda_{tc}$; in particular, we may well have
$\lambda_{tc}\sim {\cal O}(1)$ \cite{sonirev}.} We furthermore break $\lambda$ into its real and imaginary parts, $\lambda=\lambda_R+i\lambda_I$.
Also, in the framework of Model III described in \cite{sonirev},
where only one Higgs doublet acquires a VEV, one has:
\begin{eqnarray}
f^h;f^H = \cos{\tilde {\alpha}};\sin{\tilde {\alpha}}~,
\end{eqnarray}
\noindent and  
 the mixing angle $\tilde {\alpha}$ is determined by the Higgs potential.

We will also need the ${\cal H}VV$ couplings ($V=W^+,W^-$ or $Z$), 
which 
can be written in general as:
\begin{eqnarray}
{\cal L}_{{\cal H} VV} =  g_W m_W C_V c^{\cal H} {\cal H} 
g_{\mu\nu} V^{\mu} V^{\nu} \label{hww}~, 
\end{eqnarray}
\noindent where $C_W;C_Z \equiv 1;m_Z^2/m_W^2$. 
For Models II and III we have:
\begin{eqnarray}
{\rm Model~II}~& :&~ c^h;c^H=R_{11}\sin\beta
+R_{21}\cos\beta; R_{12}\sin\beta
+R_{22}\cos\beta \label{ccoupling}~,\\
{\rm Model~III}~& :&~ c^h;c^H \equiv -\sin{\tilde {\alpha}};
\cos{\tilde {\alpha}} ~.
\end{eqnarray}
\noindent Note that in the SM the only $VVh$ coupling of the one neutral 
Higgs present is $c^h=1$.
\newpage
\noindent {3. \bf $e^+e^-\to t\bar t h, t \bar t Z$; 
Cases of Tree-Level 
CP-Violation}\footnote{This section will appear in a review paper:  
"CP-Violation in Top Physics", by D. Atwood, S. Bar-Shalom, G. Eilam 
and A. Soni, to be submitted to Physics Reports.} \\

\begin{figure}
\vskip 45mm
\includegraphics{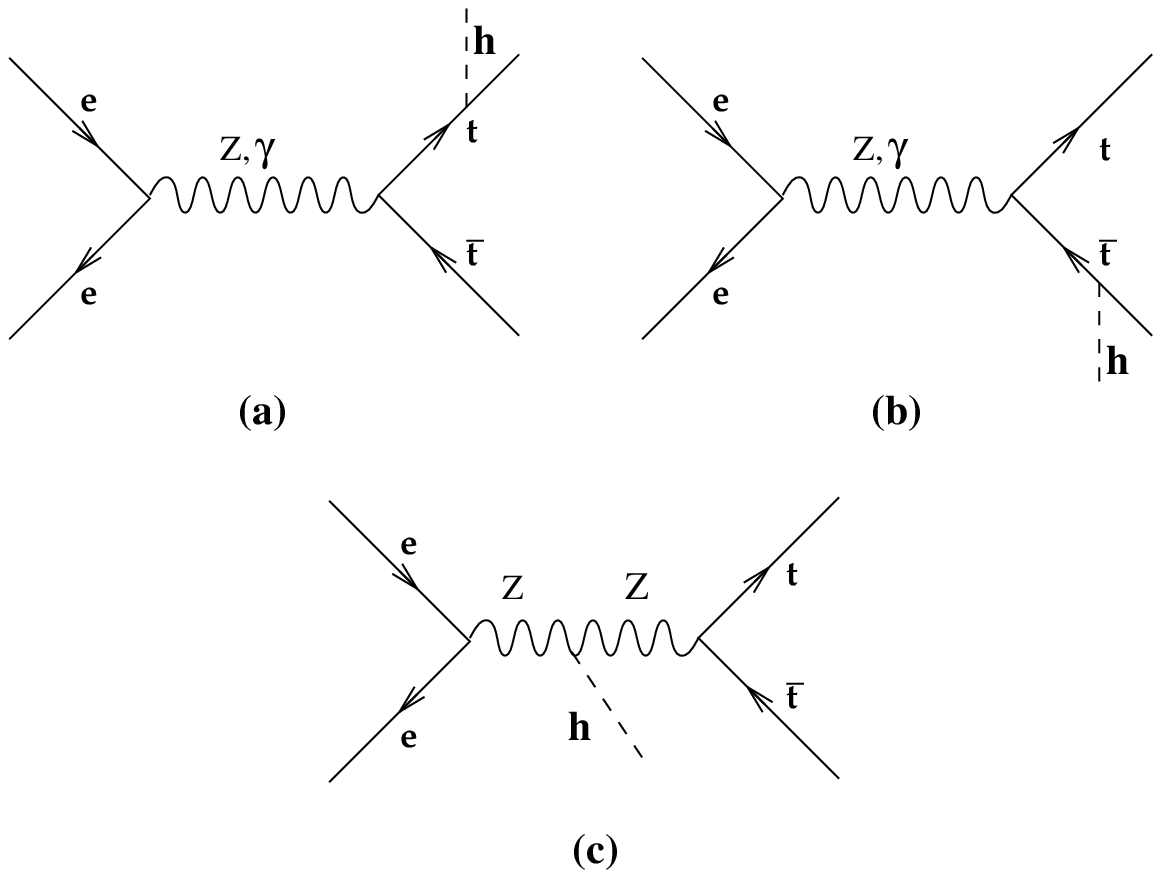}
\includegraphics{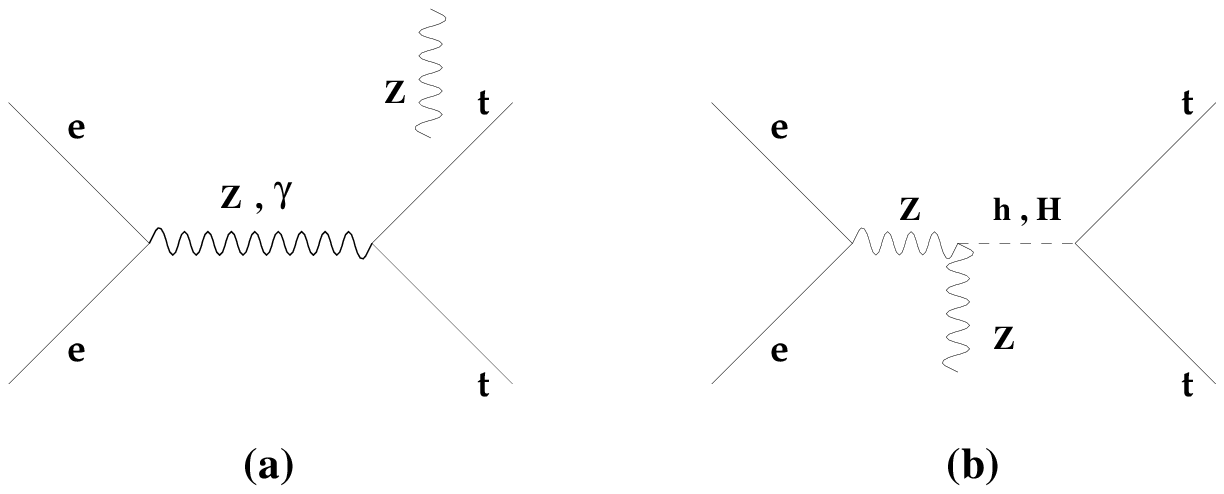}
\caption[dummy]
{Tree-level Feynman diagrams contributing to $e^+e^-\to t\bar{t} h$ 
(left hand side) and $e^+e^-\to t\bar{t} Z$ (right hand side)
in the unitary gauge, in a two Higgs doublet model. 
For $e^+e^-\to t\bar{t} Z$, 
Diagram $(a)$ on the right hand side represents 8 diagrams in which
either $Z$ or $\gamma$ are exchanged in the s-channel 
and the outgoing $Z$
is emitted from $e^+,e^-,t$ or $\bar t$.}
\label{fig1}
\end{figure}

The reactions:
\begin{eqnarray}
&& e^+(p_+)+e^-(p_-) \to t (p_t) + \bar t (p_{\bar t}) + h(p_h) 
\label{eetthzeq1} ~, \\
&& e^+(p_+)+e^-(p_-) \to t (p_t) + \bar t (p_{\bar t}) + Z(p_Z) 
\label{eetthzeq2} ~,
\end{eqnarray}
\noindent exhibits large CP violation asymmetries in a 2HDM\null. A
novel feature of these reactions is that the effect arises at tree graph
level. Basically, for the $tth$($ttZ$) final states, 
Higgs($Z$) emission off the $t$, $\bar t$ interferes with the Higgs($Z$) 
emission off the s-channel $Z$-boson (see Fig.~\ref{fig1}) 
\cite{prd53p1162,hepph9707284}. 
We find that the processes $e^+e^- \to t \bar t h$ and 
$e^+e^- \to t \bar t Z$ provide two independent, but 
analogous, promising venues to search for signatures of the 
same CP-odd phase, residing in the top-neutral Higgs 
coupling, if the value of $\tan\beta$ (the ratio 
between the two VEV's in a 2HDM) is in the vicinity of 1.
In particular, they serve 
as good examples for large CP-violating effects that could 
emanate from $t$ systems due to the large mass of the top quark 
and, thus, they 
might unveil the role of a neutral Higgs particle in  
CP-violation. 

Although these reactions are not meant (necessarily) 
to lead to the discovery of a neutral Higgs, 
they will, no doubt, be scrutinized in the 
NLC since they stand out as very interesting channels by themselves. 
In particular, they could perhaps provide a unique opportunity 
to observe the top-Higgs Yukawa couplings directly 
\cite{prl77p5172,hepph9609453,npb367p257}. 
In \cite{prl77p5172,hepph9609453}, 
using a very interesting generalization of the optimal observables 
technique used here, 
Gunion {\it et al.} have extended our work on CP-violation 
in $e^+ e^- \to t \bar t h$ described below, to include 
a detailed cross-section analysis such that all Higgs 
Yukawa couplings 
combinations are extracted. A
 detailed cross-section analysis of the reaction $e^+e^- \to t \bar t Z$
was performed in the SM by Hagiwara {\it et al.} \cite{npb367p257}.
There, it was found that the Higgs exchange contribution 
of diagram $(b)$ on 
the right hand side of 
Fig.~\ref{fig1} will be almost invisible in a 
TeV $e^+e^-$ collider for neutral Higgs
masses in the range $m_h < 2m_t$. On the contrary, we will show here that
if the scalar sector is doubled, then the lightest neutral Higgs may reveal
itself through CP-violating interactions with the top quark even if 
$m_h <2m_t$. 

In the unitary gauge the reactions in Eqs.~\ref{eetthzeq1} and \ref{eetthzeq2} 
can proceed via the Feynman diagrams depicted in Fig.~\ref{fig1}. 
We see that for $e^+ e^- \to t \bar t Z$, 
diagram $(b)$ on the right hand side of Fig.~\ref{fig1}, 
in which 
$Z$ and ${\cal H}$ are produced 
(${\cal H}=h$ or $H$ is either a real or 
virtual particle, i.e.\ $m_{\cal H} >2m_t$ or 
$m_{\cal H}<2m_t$, respectively) 
followed by ${\cal
H} \to t \bar t$, 
is the only place where new CP-nonconserving dynamics from
the Higgs  sector can arise, being proportional to the  
CP-odd phase in the
${\cal H} t \bar t$ vertex. 
As mentioned above, 
in both the $tth$ and the $ttZ$ final state 
cases, CP-violation arises due to interference of the diagrams 
where the Higgs is coupled to a Z-boson with the diagrams where 
the Higgs or $Z$ is radiated off the $t$ or $\bar t$. We note that 
in the $ttZ$ case there is no CP-violating contribution coming 
from the interference between 
the diagrams with the $ZZ{\cal H}$ coupling and the diagrams where 
the Z-boson is emitted from the incoming electron or positron lines.  
 
The relevant pieces of the interaction Lagrangian 
involve the ${\cal H} t\bar t$ and the 
${\cal H}ZZ$ couplings and is given in Eqs.~\ref{tthcoupling} 
and \ref{hww}, respectively. As   
usual the couplings $a_t^{h,H}$, $b_t^{h,H}$ and $c^{h,H}$ 
in Eqs.~\ref{mod2coup} and \ref{ccoupling}, respectively,
 are functions of
$\tan\beta \equiv v_2/v_1$ (the ratio of the two VEVs) 
and of the three mixing angles $\alpha_1$, $\alpha_2$,
$\alpha_3$ which characterize the Higgs mass matrix.
As was mentioned in the previous section, 
only two (denoted here by $h$ and $H$) 
out of the three neutral Higgs are relevant for 
the CP-violating effect studied here. 
The reason is that only two out of the three neutral Higgs particles 
in the theory can simultaneously 
have a coupling to vector bosons and a 
pseudoscalar coupling to fermions.  
We have denoted their couplings by $a_t^h,b_t^h,c^h$ and  
$a_t^H,b_t^H,c^H$, corresponding
to the light, $h$, and heavy, $H$, 
neutral Higgs, respectively. This implies 
the existence of a ``GIM-like'' cancellation, namely,
 when both $h$ and $H$ contribute
to CP-violation, then 
all CP-nonconserving effects, being proportional
to $b_t^hc^h + b_t^Hc^H$, must vanish when 
the two Higgs states $h$ and $H$
are degenerate. In the following 
we set the mass of the heavy 
Higgs, $H$, to be $m_H=750$ GeV or 1 TeV.  

In the process $e^+ e^- \to t \bar t h$, a Higgs particle is produced 
in the final state, 
therefore, the heavy Higgs-boson, $H$, is not important and 
this ``GIM-like'' mechanism is irrelevant. Note that 
there is
an additional diagram contributing to $e^+ e^- \to t \bar t h$,  
which involve the $ZhH$ coupling and 
is not shown in Fig.~\ref{fig1}. This diagram is, however, 
negligible for the large $m_H$ values used here.
In contrast, in the process 
$e^+ e^- \to t \bar t Z$, the Higgs is exchanged 
as a virtual or a real particle and the effect of $H$ is, although small compared 
to $h$, important in order to restore the ``GIM-like'' cancellation 
discussed above.  
  
For both the $tth$ and $ttZ$ final states processes, 
we denote the tree-level polarized 
differential-cross-section (DCS) by  
$\Sigma_{(j)f}$, where $f=tth$ or $f=ttZ$ corresponding 
to the $tth$ or $ttZ$ final states, respectively, and $j=1(-1)$ for 
left(right) polarized incoming electron beam. 
$\Sigma_{(j)f}$ can be subdivided 
into its CP-even ($\Sigma_{+(j)f}$) and CP-odd 
($\Sigma_{-(j)f}$) parts:
\beq
\Sigma_{(j)f} = \Sigma_{+(j)f} + 
\Sigma_{-(j)f} \label{eetthzeq4} ~.
\eeq
\noindent The CP-even and CP-odd 
DCS's can be further broken to different terms
which correspond to the various Higgs coupling 
combinations and which transform as $n$    
 under $T_N$. For both final states, $f=tth$ and $f=ttZ$, 
we have:
\begin{eqnarray}
\Sigma_{+(j)f} &=& \sum_i g_{+f}^{i(n)}
 F_{+(j)f}^{i(n)} \label{eetthzeq5} ~~,~~ {\rm CP~even} ~, \\
\Sigma_{-(j)f} &=& \sum_i g_{-f}^{i(n)}
 F_{-(j)f}^{i(n)} \label{eetthzeq6} ~~,~~{\rm CP~odd}~,
\end{eqnarray}
\noindent where $g_{+f}^{i(n)},
g_{-f}^{i(n)}$, $n=+ ~or~ -$, are different combinations  
of the Higgs couplings $a_t^{\cal H},b_t^{\cal H},c^{\cal H}$ 
and $F_{+(j)f}^{i(n)},F_{-(j)f}^{i(n)}$, 
again with $n=+~or~-$, are  kinematical functions of phase
space which transform like $n$ under $T_N$.  

Let us first write the Higgs coupling 
combinations for the CP-even part. In 
the case of $e^+ e^- \to t \bar t h$, neglecting 
the imaginary part 
in the s-channel Z propagator, we have 
four relevant coupling combinations \cite{prd53p1162,prl77p5172}:
\begin{eqnarray}
g_{+tth}^{1(+)}=(a_t^h)^2~,~g_{+tth}^{2(+)}=(b_t^h)^2~,~
g_{+tth}^{3(+)}=(c^h)^2~,~g_{+tth}^{4(+)}=a_t^h c^h \label{eetthzeq7}~.
\end{eqnarray}
\noindent In the case of $e^+ e^- \to t \bar t Z$, 
apart from the SM DCS, which corresponds to interference terms 
among the four SM diagrams represented by diagram ($a$) on 
the right hand side of 
Fig.~\ref{fig1}, and keeping terms proportional 
to both the real and imaginary parts of the 
Higgs propagator, $\Pi_{\cal H}$, we get \cite{hepph9707284}:
\begin{eqnarray}
g_{+ttZ}^{1(+)}&=&(a_t^{\cal H}c^{\cal H}) {\rm Re}(\Pi_{\cal H})~,~
g_{+ttZ}^{2(-)}=(a_t^{\cal H}c^{\cal H}) {\rm Im}(\Pi_{\cal H}) 
\label{eetthzeq8}~, \\
g_{+ttZ}^{3(+)}&=&(a_t^{\cal H}c^{\cal H})^2 {\rm Re}(\Pi_{\cal H})~,~
g_{+ttZ}^{4(+)}=(a_t^{\cal H}c^{\cal H})^2 {\rm Im}(\Pi_{\cal H})
\label{eetthzeq9}~, \\
g_{+ttZ}^{5(+)}&=&(b_t^{\cal H}c^{\cal H})^2 {\rm Re}(\Pi_{\cal H})~,~
g_{+ttZ}^{6(+)}=(b_t^{\cal H}c^{\cal H})^2 {\rm Im}(\Pi_{\cal H})
\label{eetthzeq10} ~,
\end{eqnarray}
\noindent where:
\begin{equation}
\Pi_{\cal H} \equiv \left(s+m_Z^2-m_{\cal H}^2-
2p \cdot p_Z +i m_{\cal H} \Gamma_{\cal H} \right)^{-1} 
\label{eetthzeq11}~.
\end{equation} 
\noindent $p \equiv p_-+p_+$ and $\Gamma_{\cal H}$  is the width of ${\cal
H} = h ~{\rm or}~ H$.
 
For the CP-odd parts one gets:
\begin{eqnarray}
&&g_{-tth}^{1(-)}=b_t^h c^h \label{eetthzeq12}~, \\  
&&g_{-ttZ}^{1(-)}=b_t^{\cal H} c^{\cal H} {\rm Re}(\Pi_{\cal H}) ~,~
g_{-ttZ}^{2(+)}=b_t^{\cal H} c^{\cal H} {\rm Im}(\Pi_{\cal H}) 
\label{eetthzeq13}~.
\end{eqnarray}
\begin{figure}[htb]
\psfull
 \begin{center}
  \leavevmode
  \epsfig{file=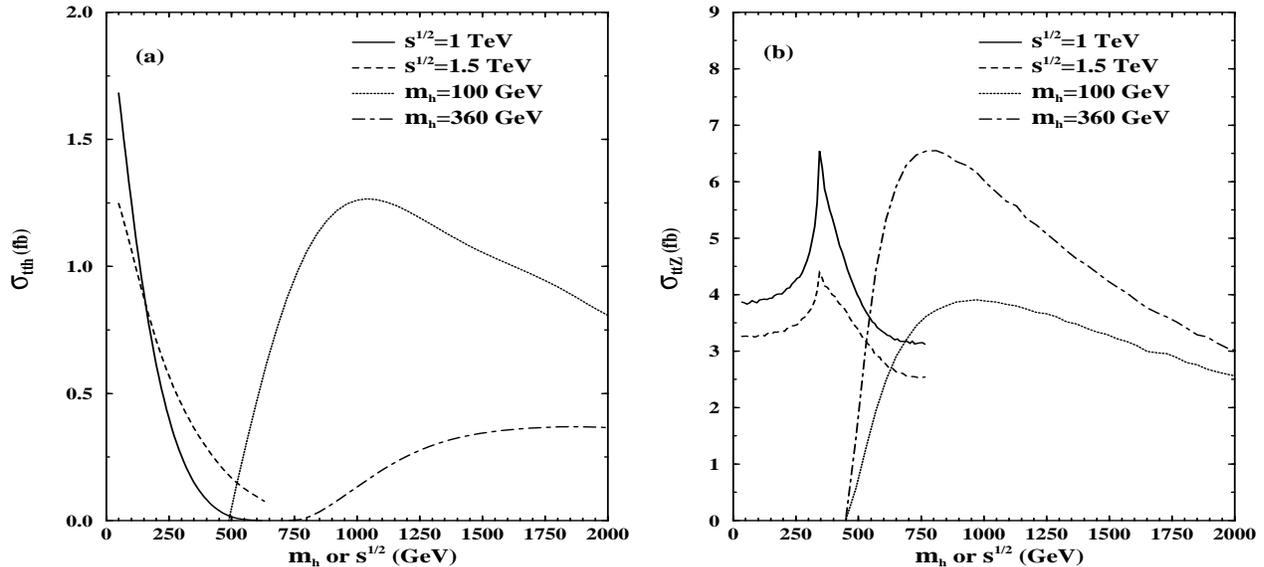,height=8cm,width=9cm,bbllx=0cm,
bblly=2cm,bburx=20cm,bbury=25cm,angle=0}
 \end{center}
\caption{The cross sections (in fb) 
for: (a) the reaction $e^+e^-\to t\bar{t}h$ with 
$\tan\beta=0.5$ and (b)
the reaction $e^+e^-\to t\bar{t}Z$ with $\tan\beta=0.3$,  
assuming unpolarized  electron and positron beams, 
for Model~II with set~II
and as a function of $m_h$ 
(solid and dashed lines) and $\sqrt s$ (dotted
and dotted-dashed lines). 
Set~II means $\left\{\alpha_1,\alpha_2,\alpha_3
\right\} \equiv \left\{\pi/2,\pi/4,0 \right\}$.} 
\label{fig2}
\end{figure}
\noindent The CP-even pieces, $\Sigma_{+(j)f}$, 
yield the corresponding 
cross-sections (recall that $f=tth$ or $ttZ$):
\beq
\sigma_{(j)f} = \int \Sigma_{+(j)f} 
(\Phi) d \Phi \label{eetthzeq14}~,
\eeq
\noindent where $\Phi$ stands for the phase-space variables. 
In Fig.~\ref{fig2}a and \ref{fig2}b 
we plot the unpolarized cross-sections, 
$\sigma_{tth}$ and $\sigma_{ttZ}$ as a function of
$m_h$ and $\sqrt s$, for Model~II,  
with $m_H=750$ GeV and the set of 
values  
$\left\{\alpha_1,\alpha_2,\alpha_3
\right\}= \left\{\pi/2,\pi/4,0 \right\}$ 
which we denote as set~II. Set~II is also adopted 
later when discussing the CP-violating effect.
Furthermore, for the $tth$ final state we choose 
$\tan\beta=0.5$ while for $ttZ$ we choose $\tan\beta=0.3$.
Afterwards, we will discuss the dependence 
of the CP-violating effect on $\tan\beta$ 
in the $tth$ and $ttZ$ cases.
One can observe the dissimilarities 
in the two cross-sections $\sigma_{tth}$ and $\sigma_{ttZ}$:
while $\sigma_{tth}$ is at most $\sim 1.5$ fb, $\sigma_{ttZ}$ 
can reach $\sim 7$ fb at around $\sqrt s \approx 750$ GeV and 
$m_h \gsim 2 m_t$. $\sigma_{tth}$ drops with $m_h$ while 
$\sigma_{ttZ}$ grows in the range $m_h \lsim 2m_t$. 
$\sigma_{ttZ}$ peaks 
at around $m_h \gsim 2m_t$ and drops as $m_h$ grows further.  
Moreover, $\sigma_{tth}$ peaks at around 
$\sqrt s \approx 1(1.5)$ TeV for $m_h=100(360)$ GeV, while 
$\sigma_{ttZ}$ peaks at around $\sqrt s \approx 750$ GeV 
for both $m_h=100$ and 360 GeV. As we will see later, these 
different features of the two cross-sections are, in part,  
the cause for the different behavior of the CP-asymmetries 
discussed below.  

Let us now concentrate on the CP-odd $T_N$-odd effects 
in $e^+ e^- \to t \bar t h; t \bar t Z$, 
emanating from the $T_N$-odd pieces in 
$\Sigma_{-(j)tth};\Sigma_{-(j)ttZ}$.  
From Eqs.~\ref{eetthzeq12} and \ref{eetthzeq13} 
it is clear that the CP-violating 
pieces
$\Sigma_{-(j)tth};\Sigma_{-(j)ttZ}$ have to be proportional 
to $b_t^h c^h$ (in the $ttZ$ case there is an additional 
similar piece corresponding to the heavy Higgs $H$). 
The corresponding CP-odd kinematic functions,  
$F_{-(j)tth}^{1(-)};F_{-(j)ttZ}^{1(-)}$, being $T_N$-odd,  
are pure tree-level quantities  
 and are proportional to the only non vanishing 
Levi-Civita tensor present, $\epsilon(p_-,p_+,p_t,p_{\bar t})$,
when the spins of the top are disregarded.  
The explicit expressions for $F_{-(j)f}^{1(-)}$ 
are:
\begin{eqnarray}
F_{-(j)tth}^{1(-)}& = & - \frac{1}{\sqrt 2}
 \left( \frac{g_W^3}{c_W^3} \right)^2 
\frac{m_t^2}{m_Z^2} \Pi_{Zh} \Pi_Z T_t^3 c_j^Z 
\epsilon(p_-,p_+,p_t,p_{\bar t}) \times
\nonumber\\ 
&& \left\{ j(\Pi_t^h + \Pi_{\bar t}^h) 
\left[ (s-s_t-m^2_h) (3w_j^- -w_j^+) + m_Z^2 (w_j^- -w_j^+) 
\right] + \right. \nonumber \\
&& \left. T_t^3 c_j^Z  \Pi_Z (\Pi_t^h-\Pi_{\bar t}^h) f \right\}
\label{eetthzeq15} ~,\\
F_{-(j)ttZ}^{1(-)}& = &
 - \sqrt 2 \left( \frac{2g_W^3}{c_W^3} \right)^2 
\frac{m_t^2}{m_Z^2} \Pi_Z T_t^3 c_j^Z \epsilon(p_-,p_+,p_t,p_{\bar t}) \times
\nonumber\\ 
&& \left\{ j (\Pi_t^Z+\Pi_{\bar t}^Z) 
\left[ m_Z^2 w_j^- + (s_t-s) w_j^+ \right] + \right. \nonumber\\
&& \left. T_t^3 c_j^Z  \Pi_Z (\Pi_t^Z-\Pi_{\bar t}^Z) f \right\} 
\label{eetthzeq16}~,
\end{eqnarray}
\noindent where
$s\equiv 2p_-\cdot p_+$ is the c.m. energy 
of the colliding electrons, $s_t\equiv (p_t+p_{\bar t})^2$ and
 $f\equiv (p_--p_+) \cdot (p_t+p_{\bar t})$. Also:
\begin{eqnarray}
\Pi_{t(\bar t)}^h & \equiv & \left(2p_{t(\bar t)} 
\cdot p_h +m^2_h \right)^{-1} ~,~ 
\Pi_{t(\bar t)}^Z \equiv \left(2p_{t(\bar t)}\cdot p_Z + m^2_Z \right)^{-1}
\label{eetthzeq17} ~, \\
\Pi_Z &\equiv& \left(s-m_Z^2\right)^{-1}~,~
\Pi_{\gamma} \equiv s^{-1}~,~
\Pi_{Zh} \equiv \left((p - p_h)^2 - m^2_Z \right)^{-1} 
\label{eetthzeq18}~.  
\end{eqnarray}
\noindent Furthermore:
\begin{equation}
w_j^{\pm} \equiv \left( s_W^2Q_t - \frac{1}{2}T_t^3 \right)
c_j^Z  \Pi_Z \pm  Q_t s_W^2c_W^2 \Pi_{\gamma} \label{eetthzeq20} ~,
\end{equation}
\noindent where $s_W(c_W)$ is the $\sin(\cos)$ of the
weak mixing angle $\theta_W$,  
$Q_f$ and $T_f^3$ are the charge and
$z$-component of the  weak isospin of 
a fermion, respectively.
$c_{-1}^Z=1/2-s_W^2, c_{1}^Z=-s_W^2$ 
(recall that $j = - 1(1)$
for a left(right)  handed electron).

Since at tree level there cannot be any absorptive phases, only
$T_N$-odd, CP asymmetries are expected to occur in 
$\Sigma_{-(j)f}$. Note that in the ttZ case 
there is a CP-odd $T_N$-even piece, 
$ b_t^{\cal H} c^{\cal H} {\rm Im}(\Pi_{\cal H}) \times 
F_{-(j)ttZ}^{2(+)}$ (see Eq.~\ref{eetthzeq13}), in the DCS. 
However, being proportional to the absorptive part 
coming from the Higgs propagator, it is not a pure 
tree-level quantity. 

Simple examples of observables that can trace 
the tree-level CP-effect in $e^+ e^- \to t \bar t h; 
t \bar t Z$ are:
\begin{equation}
O = \frac{\vec{p}_-\cdot(\vec{p}_t\times\vec{p}_{\bar{t}})}{s^{3/2}} 
 ~~,~~ O_{\rm opt} (tth;ttZ) = 
\frac{\Sigma_{-(j)tth};\Sigma_{-(j)ttZ}~
(T_N{\rm~odd~part~only})}{\Sigma_{+(j)tth};\Sigma_{+(j)ttZ}}
\label{eetthzeq21}~.
\end{equation} 
\noindent $O_{\rm opt} (tth;ttZ)$ are optimal observables 
in the sense 
that the statistical
error, in  the measured asymmetry, is minimized \cite{prd45p45}. 
As mentioned before, 
since the final state consists of three particles, using only
the available momenta, there is a unique antisymmetrical tensor that
can be formed. Thus,  
both observables are proportional  
to $\epsilon(p_-,p_+,p_t,p_{\bar t})$ and  
$O_{\rm opt} (tth;ttZ)$ are related to $O$  
through a multiplication by a CP-even function.
In the following we focus only on the CP-odd effects 
coming from the optimal observables. However, we remark 
that the results for the simple observable
$O$ exhibit the same behavior, though slightly 
smaller then those for $O_{\rm
opt}$.
 The theoretical
statistical significance, $N_{SD}$, in which an asymmetry 
can be measured
in an ideal experiment is given by 
$N_{SD}= A \sqrt L \sqrt {\sigma}$ 
($\sigma = \sigma_{tth};\sigma_{ttZ}$ 
for the $tth;ttZ$ final states), 
where for the observables $O$ and $O_{\rm opt}$, 
the CP-odd quantity $A$, defined above, is: 
\begin{equation}
A_O \approx \langle O \rangle/\sqrt{\langle O^2\rangle} ~~,~~
A_{\rm opt} \approx \sqrt{\langle O_{\rm opt} \rangle} 
\label{eetthzeq22} ~.
\end{equation}
\noindent Also, $L$ is the effective luminosity 
for fully reconstructed
$t \bar t h$ or $t \bar t Z$ events. 
In particular,  we take $L=\epsilon {\cal L}$,
where ${\cal L}$ is the total yearly integrated  
luminosity and $\epsilon$ is the overall efficiency for   
reconstruction of the $t \bar t h$ or $t \bar t Z$ final states. 

\begin{figure}[htb]
\psfull
 \begin{center}
  \leavevmode
  \epsfig{file=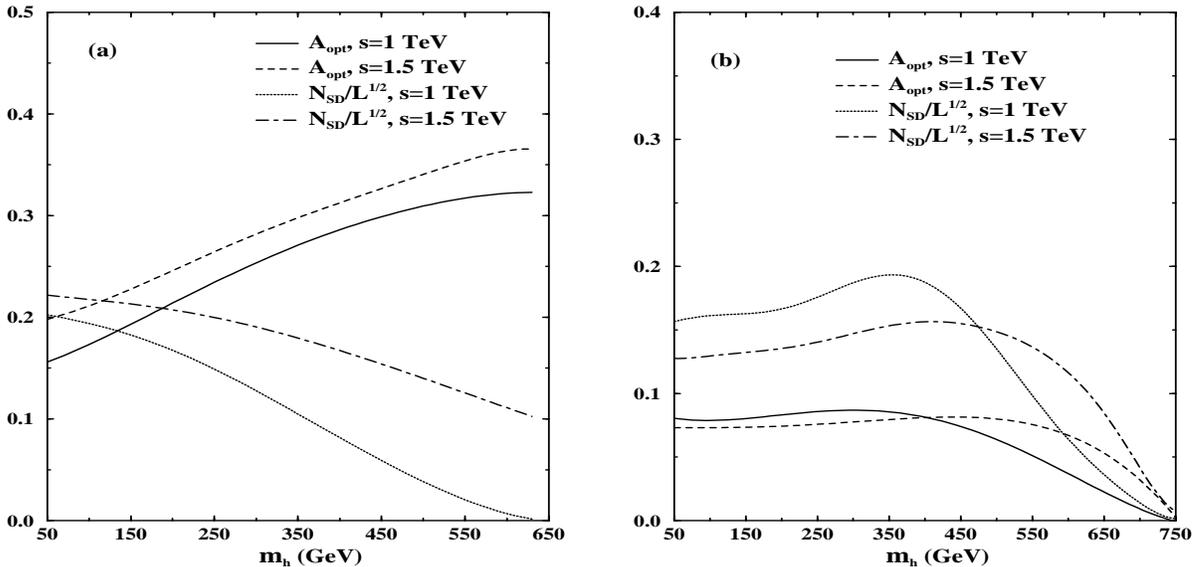,height=8cm,width=9cm,bbllx=0cm,
bblly=2cm,bburx=20cm,bbury=25cm,angle=0}
 \end{center}
\caption{The asymmetry, $A_{\rm opt}$, and scaled statistical 
significance, $N_{SD}/\sqrt L$, for the optimal observable 
$O_{\rm opt}$ for: (a) the reaction $e^+e^-\to t\bar{t}h$ with 
$\tan\beta=0.5$ and (b)
the reaction $e^+e^-\to t\bar{t}Z$ with $\tan\beta=0.3$,
as a function of the light Higgs mass $m_h$, for
$\sqrt s=1$ TeV and 1.5 TeV\null. All graphs are with set~II 
of the parameters, as in Fig.~\ref{fig2}.}
\label{fig3}
\end{figure}

For numerical results 
we have used set~II defined above for the 
angles $\alpha_{1,2,3}$, i.e.   
$\left\{\alpha_1,\alpha_2,\alpha_3
\right\}= \left\{\pi/2,\pi/4,0 \right\}$ (recall that for 
the $tth$ final state we choose $\tan\beta=0.5$ while for 
the $ttZ$ final state we choose $\tan\beta=0.3$). 
Figs.~\ref{fig3}a and \ref{fig3}b show 
the expected asymmetry and statistical significance in the unpolarized case,
corresponding to 
$O_{\rm opt}$ in Model~II for the $tth$ and $ttZ$ final states, 
respectively. The asymmetry is plotted 
as a function of the mass of the light 
Higgs ($m_h$) where again, $m_H=750$ GeV in the $ttZ$ case. 
We plot $N_{SD}/\sqrt L$,
thus scaling out the luminosity factor from the 
theoretical prediction (as a reference value, we 
note that for $L = 100$ fb$^{-1}$, 
$N_{SD}/\sqrt L =0.1$ will correspond to a one-sigma effect).
We remark that set~II corresponds to the largest CP-effect, though not
unique. In the $tth$ case $\tan\beta=0.5$ is favored, however, 
the effect mildly depends on $\tan\beta$ in the range 
$ 0.3 \lsim \tan\beta \lsim 1$ (see also \cite{prd53p1162}).  
In the $ttZ$ case, the effect is practically 
insensitive to $\alpha_3$ and is roughly proportional to $1/\tan\beta$, 
it therefore drops as $\tan\beta$ is increased. 
Nonetheless, we find that $N_{SD}/\sqrt L>0.1$, even in the unpolarized 
case for $\tan\beta \lsim 0.6$ (see also \cite{hepph9707284}). 

\begin{figure}[htb]
\psfull
 \begin{center}
  \leavevmode
  \epsfig{file=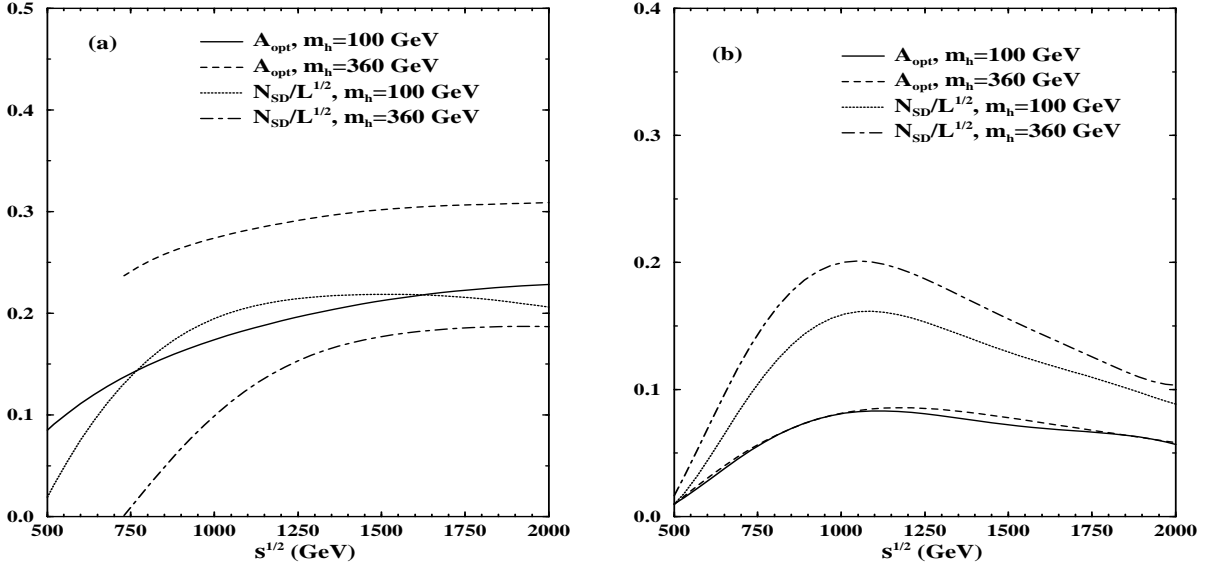,height=8cm,width=9cm,bbllx=0cm,
bblly=2cm,bburx=20cm,bbury=25cm,angle=0}
 \end{center}
\caption{The asymmetry, $A_{\rm opt}$, and scaled statistical 
significance, $N_{SD}/\sqrt L$, for the optimal observable 
$O_{\rm opt}$ for: (a) the reaction $e^+e^-\to t\bar{t}h$ with 
$\tan\beta=0.5$ and (b)
the reaction $e^+e^-\to t\bar{t}Z$ with $\tan\beta=0.3$,
as a function of the c.m. energy $\sqrt s$, for 
$m_h=100$ GeV and $m_h=360$ GeV. 
All graphs are with set~II 
of the parameters, as in Fig.~\ref{fig2}.}
\label{fig4}
\end{figure}

From Fig.~\ref{fig3}a we see that, in the $tth$ case, 
as $m_h$ grows the asymmetry increases while the statistical 
significance drops, in part because of the decrease in the cross-section. 
We can see that the asymmetry can become extremely large and it 
ranges from $\sim 15\%$, for a Higgs mass below 100 GeV, to 
$\sim 35\%$ for $m_h \sim 600$ GeV. 
Evidently, the CP-effect is more significant for smaller masses 
of $h$, whereas $A_{\rm opt}$ is smaller. 
In contrast, from Fig.~\ref{fig3}b we see that, 
in the $ttZ$ case, $A_{\rm opt}$ stays roughly 
fixed at around $7-8 \%$, for $m_h \lsim 2m_t$, 
and then drops till it totally vanishes 
at $m_h=m_H=750$ GeV, in which case the ``GIM-like'' 
mechanism discussed
before applies. The scaled statistical significance 
$N_{SD}/\sqrt L$ behaves roughly as $A_{\rm opt}$. That is,    
$N_{SD}/\sqrt L \approx 0.12-0.2$ in the mass range 
$50 ~{\rm GeV} \lsim m_h \lsim 350 ~{\rm GeV}$,   
for both $\sqrt s=1$ and 1.5 TeV.   

Figs.~\ref{fig4}a and \ref{fig4}b show the 
dependence of $A_{\rm opt}$ and $N_{SD}/\sqrt L$ on the 
c.m. energy, $\sqrt s$, for the $tth$ and $ttZ$ cases, respectively.
We see that, in the case of $tth$, the CP effect peaks at
$\sqrt s \approx 1.1(1.5)$ TeV for $m_h=100(360)$ GeV and 
stays roughly the same as $\sqrt s$ is further increased to 2 TeV. 
In the case of $ttZ$, the statistical significance is maximal 
at around $\sqrt s \approx 1$ TeV and then 
decreases as $\sqrt s$ grows, for both $m_h=100$ and 360 GeV. 
Contrary to the $tth$ case, where a light $h$ is favored, in the 
$ttZ$ case, the effect is best for $m_h \gsim 2m_t$. In that 
range,
on-shell $Z$ and $h$ are produced followed by the $h$ decay 
$h \to t \bar t$, thus, the 
Higgs exchange diagram becomes more dominant.           

In Tables~\ref{eetthztab1} and \ref{eetthztab2} we 
present $N_{SD}$ for $O_{\rm opt}$, for the $tth$ and $ttZ$ cases, 
respectively, in Model~II with set~II,
and we also compare 
the effect of beam polarization with the unpolarized case.
As before, we take $\tan\beta=0.5$ and $\tan\beta=0.3$ 
for the $tth$ and $ttZ$ cases, respectively, 
where for the $ttZ$ case we also present numbers for both 
$m_H=750$ GeV (shown in the parentheses) and $m_H=1$ TeV, to
demonstrate the sensitivity of the 
CP-effect to the mass of the 
heavy Higgs.
For illustrative 
purposes, we choose
$m_h=100,160$ and 360 GeV and show the 
numbers for $\sqrt s =1$ TeV 
with ${\cal L}=200$ [fb]$^{-1}$ and for 
$\sqrt s =1.5$ TeV with ${\cal L}=500$
[fb]$^{-1}$ (see \cite{nlc}). 
In both cases we take $\epsilon=0.5$ assuming that there is
no loss  of luminosity when the electrons are polarized 
(if the efficiency for $t \bar th$ and/or $ttZ$  
reconstruction is $\epsilon=0.25$, then our numbers 
would correspondingly require 2 years of running). 

Evidently, for both reactions, 
left polarized incoming electrons can probe CP-violation 
slightly better
than unpolarized ones.
We can see that in the $tth$ case the CP-effect 
drops as the mass of the light Higgs $h$ grows, while 
in the $ttZ$ case it grows with $m_h$. In particular, we find 
that with $\sqrt s =1.5$ TeV and 
for $m_h \gsim 2m_t$ the effect is comparable for both 
the $tth$ and the $ttZ$ cases where it reaches above $3 \sigma$ 
for negatively polarized electrons. With a light Higgs mass 
in the range $100 ~{\rm GeV} \lsim m_h \lsim 160 ~{\rm GeV}$, 
the $tth$ case is more sensitive to $O_{\rm opt}$ and the 
CP-violating effect 
can reach $\sim 4 \sigma$ for left polarized electrons. 
In that light Higgs mass range, the CP-violating effect reaches 
slightly below $2.5 \sigma$ for the $ttZ$ case. 
For a c.m. energy of $\sqrt s =1$ TeV and $m_h =360$ GeV, 
the $ttZ$ case is much more sensitive to $O_{\rm opt}$ and 
the effect can reach $2.2 \sigma$ for left polarized electron beam. 
However, with that c.m. energy, the $tth$ mode gives a better 
CP-odd effect in the range 
$100 ~{\rm GeV} \lsim m_h \lsim 160 ~{\rm GeV}$.
\begin{table}
\begin{center}
\caption[first entry]{The statistical significance, $N_{SD}$, in which 
the CP-nonconserving effects in $e^+e^- \to t \bar t h$ can be detected in
one year of running of a future high energy  collider with either unpolarized
or polarized incoming electron beam.  
We have used $\tan\beta=0.5$, a yearly integrated luminosity
of ${\cal L}=200$ and 500 [fb]$^{-1}$ for 
$\sqrt s=1$ and 1.5 TeV, respectively,
and an efficiency reconstruction factor of  $\epsilon=0.5$ 
for both energies. 
Recall that $j=1(-1)$ stands  for right(left) polarized electrons. 
Set~II means $\left\{\alpha_1,\alpha_2,\alpha_3
\right\} \equiv \left\{\pi/2,\pi/4,0 \right\}$.

\bigskip

\protect\label{eetthztab1}}
\begin{tabular}{|r||r|r|r|r|} \cline{3-5}
\multicolumn{2}{c||}{~~} & \multicolumn{3}{c|}{$e^+e^- \to t \bar t h$ (Model~II with Set~II)}\\ \hline
$\sqrt{s}$ & j
& \multicolumn{3}{c|}{$O_{\rm opt}$}
\\ \cline{3-5}
$({\rm TeV}) \Downarrow$ &$({\rm GeV}) \Rightarrow$ & $m_h=100$ & $m_h=160$ &$m_h=360$\\ 
\hline
\hline
&-1& $2.2$ & $2.0$ & 1.1 \\ \cline{2-5}
1 & unpol & $2.0$ & $1.9$ & 1.0\\ \cline{2-5}
& 1 &  $1.8$ & $1.7$ & 0.9 \\ \hline
\hline
& -1 &  $4.0$ & $3.9$ & 3.2 \\ \cline{2-5}
1.5 & unpol & $3.6$ & $3.5$ & 2.9 \\ \cline{2-5}
&1 &  $3.2$ & $3.1$ & 2.6\\ \hline 
\end{tabular}
\end{center}
\end{table}

\begin{table}
\begin{center}
\caption[first entry]{The same as Table~\ref{eetthztab1} but for 
$e^+e^- \to t \bar t Z$, with $\tan\beta=0.3$. 
In this reaction, effects of the heavy 
Higgs, $H$, are included and  
$N_{SD}$ is given for both $m_H=750$ GeV (in parentheses) 
and $m_H=1$ TeV.

\bigskip

\protect\label{eetthztab2}}
\begin{tabular}{|r||r|r|r|r|} \cline{3-5}
\multicolumn{2}{c||}{~~} & \multicolumn{3}{c|}{$e^+e^- \to t \bar t Z$ (Model~II with Set~II)}\\ \hline
$\sqrt{s}$ & j
& \multicolumn{3}{c|}{$O_{\rm opt}$}
\\ \cline{3-5}
$({\rm TeV}) \Downarrow$ &$({\rm GeV}) \Rightarrow$ & $m_h=100$ & $m_h=160$ &$m_h=360$\\ 
\hline
\hline
&-1& $(1.8)~1.7$ & $(1.8)~1.8$ & (2.2)~2.2 \\ \cline{2-5}
1 & unpol & $(1.6)~1.6$ & $(1.7)~1.6$ & (2.0)~2.0 \\ \cline{2-5}
& 1 &  $(1.5)~1.5$ & $(1.5)~1.5$ & (1.8)~1.8 \\ \hline
\hline
& -1 &  $(2.3)~2.9$ & $(2.4)~3.0$ & (2.8)~3.3 \\ \cline{2-5}
1.5 & unpol & $(2.1)~2.6$ & $(2.1)~2.7$ & (2.5)~3.0 \\ \cline{2-5}
&1 &  $(1.8)~2.3$ & $(1.8)~2.3$ & (2.1)~2.6\\ \hline 
\end{tabular}
\end{center}
\end{table}
Before continuing, let us summarize the above results and add 
some concluding remarks. We have shown that an extremely interesting 
CP-odd signal may arise at tree-level in the reactions 
$e^+ e^- \to t \bar t h$ and $e^+ e^- \to t \bar t Z$.
The asymmetries that were found are $\sim 15\% - 35 \%$ in the 
$tth$ case and $\sim 7\% - 8 \%$ for the $ttZ$ final state. 
These asymmetries can give rise to a $\sim 3-4 \sigma$ 
CP-odd signals in a future $e^+e^-$ collider running with c.m. 
energies in the range 
$1 ~{\rm TeV} \lsim \sqrt s \lsim 2 ~{\rm TeV}$. 

Note, however, that the simple observable, $O$, as well as 
the optimal one, $O_{\rm opt}$, 
require the identification of the $t$ and $\bar t$ and the 
knowledge of
the transverse components of their momenta in each
$t\bar th$ or $t \bar t Z$ event. 
Thus, for the main top decay, $t \to b W$,
 the most suitable scenario is when either the $t$ or the $\bar
t$ decays  semi-leptonically and the other decays hadronically. 
Distinguishing
between $t$ and $\bar t$ in the double hadronic 
decay case will require 
more effort and still remains an experimental challenge.
Note, for example, 
that if the identification of the charge of the b-jets coming from 
the $t$ and the $\bar t$ is possible then the difficulty 
in reconstructing 
the transverse components of the $t$ and $\bar t$ 
momenta
can be bypassed by using the momenta of the decay products
 in the
processes $e^+e^-\to t\bar th\to bW^+\bar bW^- h$ 
and $e^+e^-\to t\bar tZ\to bW^+\bar bW^- Z$. For example, the
observable:  
\beq 
O_b = \frac{\epsilon (p_-, p_+, p_b, p_{\bar b}) } 
{s^2} \label{eetthzeq23} ~,
\eeq 
\noindent can be constructed. 
We have considered this observable for the reaction  
$e^+e^-\to t\bar th\to bW^+\bar bW^- h$ in \cite{prd53p1162}.
We found there that, close to threshold, this observable is not
very effective. However at higher energies,  
$O_b$ is
about as sensitive as the simple triple product 
correlation $O$ defined in Eq.~\ref{eetthzeq21} and, therefore,
slightly less sensitive then $O_{\rm opt}$.

Note also that for the light Higgs mass, $m_h=100$ GeV, 
the most suitable way to detect the Higgs 
in $e^+e^-\to t\bar th \to bW^+\bar bW^- h $ 
is via $h \to b \bar b$ 
with branching ratio $\sim 1$.   
For $m_h \gsim 2m_t$, and specifically with set~II used above, 
there are two competing Higgs decays, $h \to t \bar t$ and 
$h \to W^+ W^-$, depending on the value of 
$\tan\beta$. For example, for $\tan\beta=0.5$, as was chosen above, 
one has ${\rm Br} (h \to t \bar t) \approx 0.77$ and 
${\rm Br} (h \to W^+ W^-) \approx 0.17$, thus, the 
$h \to t \bar t$ mode is more adequate. Of course, 
$h \to t \bar t$ will dominate more for smaller values of 
$\tan\beta$ and less if $\tan\beta > 0.5$. In particular, 
for $\tan\beta=0.3(1)$ one has 
${\rm Br} (h \to t \bar t) \approx 0.89(0.57)$ and 
${\rm Br} (h \to W^+ W^-) \approx 0.08(0.32)$. 
 
As emphasized before, the final states 
$t\bar th$ and $t\bar tZ$, and in particular the $t\bar th$,
 are expected to be the center of
considerable attention at a future linear collider. 
Extensive studies
of these reactions 
 are expected to teach us about the details of the
couplings of Higgs to the top quark. Thus, it is gratifying that the
same final states promise to exhibit interesting effects of CP
violation. It would be very instructive to examine the effects in other
extended models. Numbers emerging from the 2HDM that was used
especially with the specific value of the parameters, should 
be viewed as an illustrative example.
The important point is that the
reactions $e^+e^-\to t\bar th \to bW^+\bar bW^- h $ 
and $e^+e^-\to t\bar tZ \to bW^+\bar bW^- Z$
appear to be
very powerful and very clean tools 
for extracting valuable information on
the parameters of the underlying model for CP violation.\\

\noindent {3. \bf $e^+ e^- \to t \bar c \nu_e {\bar {\nu}}_e,~ 
t \bar c e^+ e^-$; Cases of Tree-Level 
Flavor-Changing-Scalar Transitions} \\

\begin{figure}[htb]
\psfull
 \begin{center}
  \leavevmode
  \epsfig{file=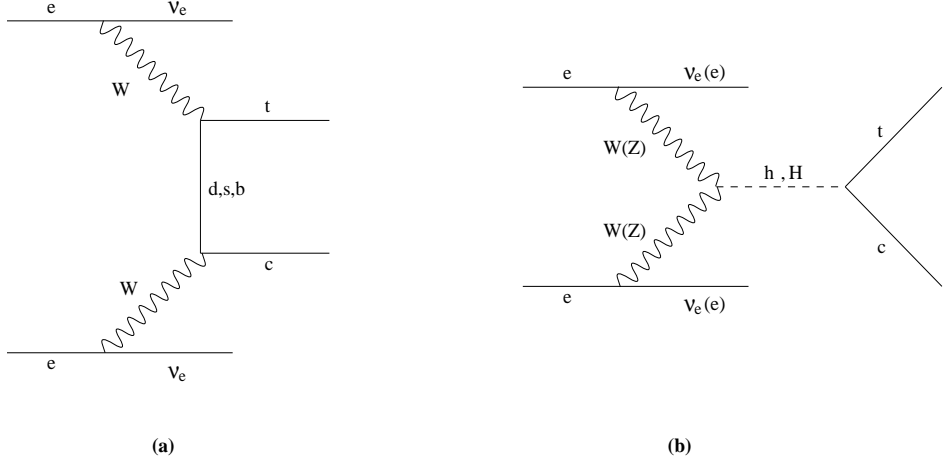,height=8cm,width=9cm,bbllx=0cm,
bblly=2cm,bburx=20cm,bbury=25cm,angle=0}
 \end{center}
\caption{(a) The Standard Model diagram for $e^+e^- \to t \bar c \nu_e
{\bar {\nu}}_e$; (b) Diagrams for $e^+e^- \to t \bar c \nu_e {\bar {\nu}}_e
(e^+e^-)$ in Model~III\null.}
\label{fig5}
\end{figure}

The reactions:
\begin{eqnarray}
e^+e^- \to t \bar c \nu_e {\bar \nu_e};~ \bar t c \nu_e {\bar \nu_e}~,~
e^+e^- \to t \bar c e^+ e^-;~\bar t c e^+ e^- \label{eetcnn}~,
\end{eqnarray}
\noindent occur via $W^+ W^-$ or $ZZ$ fusion (see Fig.~\ref{fig5}). The  
FCS transitions in those reactions gives rise to appreciable 
cross-sections, at the level of few fb's \cite{prl79p1217}, 
which should
be accessible to the Next generation of $e^+e^-$ Linear
Colliders.

The crucial interesting feature of the $VV$ 
fusion reactions is that, being
a t-channel fusion process, the corresponding cross-sections 
{\it grow} with
the c.m.\ energy of the collider. Therefore, 
even if no $t \bar c$ events are detected at
$\sqrt s =500$ GeV via the previously proposed processes such as 
$e^+e^- \to t \bar c;~t \bar c f \bar f;~t \bar t
c \bar c$ (see Atwood {\it et al.} 
and Hou {\it et al.} in \cite{fc2hdm}), 
there is still a strong motivation to look for a 
signature of the $VV$ fusion processes in Eq.~\ref{eetcnn},
 especially at somewhat higher energies.

We note at this point that in the SM, the parton level reaction 
$W^+ W^- \to t \bar c$ can also proceed at tree-level via diagram (a)
in Fig.~\ref{fig5}. However, numerically, the corresponding cross-section,
$\sigma^{\nu \nu t c}_{SM} = 
\sigma_{SM} (e^+e^- \to t \bar c \nu_e {\bar \nu_e} + 
\bar t c \nu_e {\bar \nu_e})$, is found to be of no experimental 
relevance. In particular, we found that $\sigma^{\nu \nu
tc}_{\rm SM} \approx 10^{-5}-10^{-4}$ fb 
for $\sqrt s = 0.5 - 2$ TeV due to a severe CKM 
suppression \cite{prl79p1217}. 
We therefore neglect the SM contribution in the following.   

In Model~III, where the $tc{\cal H}$ coupling of Eq.~\ref{htc} 
are present, 
$VV \to t \bar c$ proceeds at tree-level
via the $\hat s$-channel neutral Higgs exchange 
of diagram (b) in Fig.~\ref{fig5}. 
Neglecting the SM diagram, 
the corresponding parton-level cross-section
${\hat \sigma}_V \equiv {\hat \sigma}
(V^1_{\lambda^1}V^2_{\lambda^2} \to t \bar c)$
is given by~\cite{prl79p1217}:\footnote{$V^1=W^+$,$V^2=W^-$ for 
$W^+W^-$ fusion and $V^1=V^2=Z$ for $ZZ$ fusion.}
\begin{eqnarray}
{\hat \sigma}_V &=
\frac{\left( \sin 2{\tilde \alpha } \right)^2 N_c \pi \alpha^2}{4 {\hat s}^2
\beta_V s^4_W}  \left(\frac{m_V}{m_W}\right)^4 
| \epsilon^{V^1}_{\lambda^1} \cdot \epsilon^{V^2}_{\lambda^2}|^2  
|\Pi_h - \Pi_H |^2 \times 
\nonumber \\
&  m_t m_c \sqrt {a_+ a_-} ( a_+ \lambda_R^2 + a_- \lambda_I^2 )    
\label{vvhtc}~,
\end{eqnarray}
\noindent where $\epsilon^{V^i}_{\lambda^i}$ ($i=1,2$) 
is the polarization vector of $V^i$ with helicity $\lambda^i$. Also:
\begin{eqnarray}
a_\pm = \hat s - ( m_t \pm m_c )^2~~,~~ 
\beta_\ell \equiv \sqrt {1-4 m_{\ell}^2/\hat s}~,
\end{eqnarray}
\noindent and:
\begin{eqnarray}
\Pi_{\cal H} = \frac{1}{\left( \hat s - m_{\cal H}^2 +i m_{\cal H} 
{\Gamma_{\cal H}} \right)}~.
\end{eqnarray}
For definiteness, we will ignore CP violation and take 
$\lambda_I=0$ and
$\lambda=\lambda_R$ in Eq.~\ref{htc}.
In calculating the full cross sections, 
i.e. $\sigma^{\nu \nu t c} \equiv 
\sigma (e^+e^- \to t \bar c \nu_e {\bar \nu_e} + \bar t c \nu_e {\bar \nu_e})$ and $\sigma^{e e t c} \equiv
\sigma (e^+e^- \to t \bar c e^+ e^- + \bar t c e^+ e^-)$, 
we used the 
effective vector boson approximation (EVBA)
\cite{cahn}. An exact calculation of $\sigma^{\nu \nu t c}$, using 
$2 \to 4$ helicity amplitudes, was performed in \cite{hou}, 
where it was found that, in the ranges 
where $\sigma^{\nu \nu t c} \gsim 1$ fb, the difference between the EVBA 
and the exact calculation is about 
$\sim 10\%$. Note also from Eq.~\ref{vvhtc}, that 
${\hat \sigma}_W \to {\hat \sigma}_Z$
for $m_W \to m_Z$. The main difference between 
$\sigma^{\nu \nu tc}$ and $\sigma^{eetc}$
then arises from the dissimilarity between the distribution functions
for $W$ and $Z$ bosons, and we find that 
$\sigma^{\nu \nu tc} \approx 10 \times \sigma^{eetc}$ (for more details 
see \cite{prl79p1217}). Therefore, below we 
present an analysis of 
 $\sigma^{\nu \nu tc}$ only, keeping in mind
that $\sigma^{eetc}$ exhibits the same behavior though suppressed by about 
an order of magnitude.  

As mentioned in the introduction, only two out of the three 
neutral Higgs particles are relevant for the present 
analysis. The reason is that, in the FC case also, only 
$h$ and $H$ can simultaneously have a coupling to a vector 
boson and a FC coupling to $t \bar c$. 
therefore there is a ``GIM-like'' cancellation in the scalar 
sector, operative also 
in the flavor-changing effects of Model~III. In particular, 
the choice  $\tilde {\alpha}= \pi/4$, for which 
the $tch$ and $tcH$ couplings are identical (see Eq.~\ref{htc}), 
is special in the sense that for
this value the ``GIM-like'' 
cancellation mentioned above is most effective. Thus, for 
degenerate $h$ and $H$ masses and with $\tilde {\alpha}= \pi/4$, 
the cross-section $\sigma^{\nu \nu tc}$ vanishes. 
However, for $\tilde {\alpha} \neq \pi/4$, this 
``GIM-like'' cancellation is only partly effective and 
$\sigma^{\nu \nu tc} \gsim 1$ fb is still possible, even 
with $m_H=m_h$. 

\begin{figure}[htb]
\psfull
 \begin{center}
  \leavevmode
  \epsfig{file=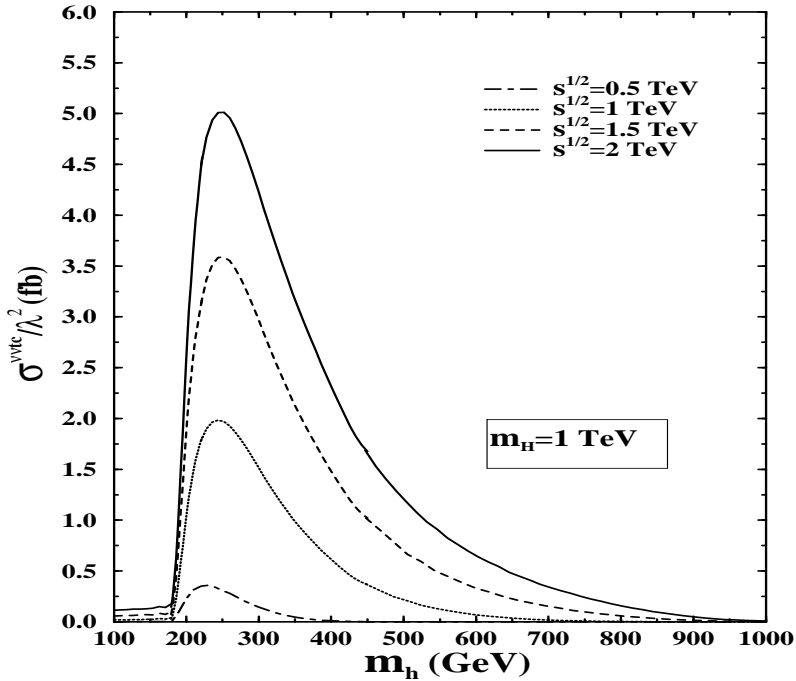,height=8cm,width=9cm,bbllx=0cm,
bblly=2cm,bburx=20cm,bbury=25cm,angle=0}
 \end{center}
\caption{The cross-section $\sigma (e^+e^- \to t \bar c \nu_e {\bar
{\nu}}_e +  \bar t c \nu_e {\bar {\nu}}_e)$ in units of $\lambda^2$
as a function of $m_h$ for $\sqrt s=0.5,1,1.5$ and 2 TeV\null. $\tilde\alpha=\pi/4$ and we have set $\lambda=1$ in the width $\Gamma_{\cal H}$.}
\label{fig6}
\end{figure}

In Fig.~\ref{fig6} we show the dependence of the 
scaled cross-section
$\sigma^{\nu \nu tc}/\lambda^2$ on the mass of the 
light Higgs $m_h$ for four values of $s$ and for    
$\tilde {\alpha}= \pi/4$.\footnote{The scaled cross-section,
$\sigma^{\nu \nu tc}/ \lambda^2$, has a residual mild dependence on
$\lambda$ through its dependence on the Higgs particles 
widths, $\Gamma_{\cal H}$.}
The cross-section peaks at
$m_h \simeq 250$ GeV and drops as the mass of the light Higgs approaches
that of the heavy Higgs due to the ``GIM-like'' 
cancellation discussed above. 
For c.m. energies of $\sqrt s \gsim 1$ TeV,  
$\sigma^{\nu \nu tc} \gsim 2$ fb at the point 
$\tilde {\alpha}= \pi/4$, if  
$\lambda=1$ and $m_h \approx 250$ GeV\null.
It is therefore evident from Fig.~\ref{fig6} that 
at an NLC running at energies of
$\sqrt s \gsim 1$ TeV and an integrated luminosity of the order
of ${\cal L} \gsim 10^2$ [fb]$^{-1}$, Model~III (with $\lambda=1$) 
predicts
hundreds and
up to thousands of $t \bar c \nu_e {\bar {\nu}}_e$ events and 
several tens
to hundreds of $t \bar c e^+ e^-$ events. For
example, with $\sqrt s=1.5$ TeV, ${\cal L}= 500$ [fb]$^{-1}$~\cite{nlc}, and
$m_h \approx 250$ GeV, $\tilde\alpha =\pi/4$, $\lambda=1$, the cross-section
$\sigma^{\nu \nu tc}$($\sigma^{eetc}$) would yield about 2000(200)
such events. Note also that even with $m_h \approx 500$ GeV,
this projected luminosity will still yield hundreds of
$t \bar c \nu_e {\bar {\nu}}_e$ events and tens of
$t \bar c e^+ e^-$ events at $\sqrt s=1.5$ TeV. 
The corresponding SM prediction
yields, as mentioned above, essentially zero events.

\begin{figure}[htb]
\psfull
 \begin{center}
  \leavevmode
  \epsfig{file=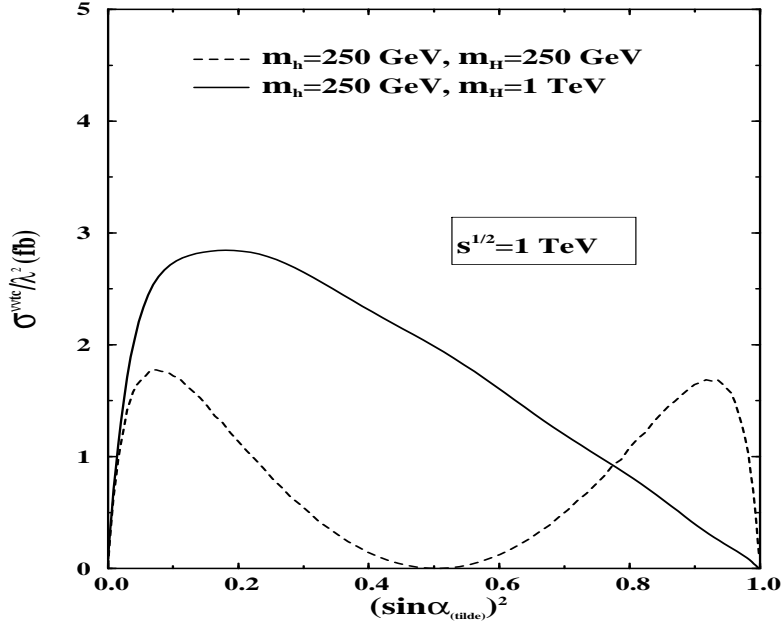,height=8cm,width=9cm,bbllx=0cm,
bblly=2cm,bburx=20cm,bbury=25cm,angle=0}
 \end{center}
\caption{The cross-section $\sigma (e^+e^- \to t \bar c \nu_e {\bar
{\nu}}_e +  \bar t c \nu_e {\bar {\nu}}_e)$ in units of $\lambda^2$
as a function of $(\sin \tilde {\alpha})^2$ for $\sqrt s=1$ TeV, $m_h=250$
GeV and $m_H=250,1000$ GeV\null. $\lambda$ as in Fig.~\ref{fig6}.}
\label{fig7}
\end{figure}

In Fig.~\ref{fig7} we show the dependence of 
$\sigma^{\nu \nu tc}/\lambda^2$ on 
$(\sin\tilde {\alpha})^2$ for $m_h=250$
GeV, $\sqrt s = 1$ TeV and for two possible values of $m_H$, $m_H=250$ GeV
and $m_H=1$ TeV\null. The same behavior is observed for any value of 
$\sqrt s$ in
the range 0.5--2 TeV\null.
We see that in the large splitting case, i.e. $m_H=1$ TeV,  
$\sigma^{\nu \nu tc}( \pi/ 14 \lsim \tilde {\alpha} \lsim \pi/4) >
\sigma^{\nu \nu tc}(\tilde {\alpha}= \pi/4)$. Moreover, even for $(m_H-m_h)
= 0$, $\sigma^{\nu \nu tc} \gsim 1$ fb is still possible for $ 0.02 \lsim
(\sin\tilde {\alpha})^2 \lsim 0.22$ and $ 0.78 \lsim (\sin\tilde {\alpha})^2
\lsim 0.98$. In fact, our analysis shows that, with moderate restrictions on
$\tilde {\alpha}$, $\sigma^{\nu \nu tc}$
remains well above the fb level for $\sqrt s \gsim 1$ TeV
as long as one of the neutral Higgs particles 
(either the light one or the heavy one) is kept within
$200 ~{\rm GeV} \lsim m_{\cal H} \lsim 400~{\rm GeV}$, while the mass of the
other
Higgs can take practically any value between 100 GeV -- 1000 GeV\null.
As was pointed out in \cite{hou}, it is interesting to 
note that if both $m_h$ and $m_H$ are of order of the weak scale
and the splitting between the two masses is a few times the width of 
both Higgs bosons, then, with $m_h \approx 250$ GeV, $\sigma^{\nu \nu tc}$ 
can reach $\sim 8$ fb at $\sqrt s =2$ TeV. This is almost
 twice as large as the $m_H=1$ TeV case.   
Also, Hou {\it et al.} have shown that the cross-section 
$\sigma^{\nu \nu tc}$ depends only mildly on $\tilde\alpha$, when 
$m_h$ and $m_H$ are not degenerate with values 
of the order of few hundreds GeV. In particular, they 
showed that the dependence on $\tilde\alpha$ is only significant 
near the edge values $\tilde\alpha=0,\pi/2$, for which 
the $W^+W^- \to t \bar c$ amplitude vanishes.

To summarize this section,
it was shown that the cross-section 
for the reaction $e^+ e^- \to t \bar c \nu_e {\bar\nu}_e +
\bar t c \nu_e {\bar\nu}_e$ can reach a few fb's in a rather wide 
range of the Model~III parameter space.\footnote{Recall 
that the cross-section is 
$\propto \lambda^2$ so that even a moderate change of $\lambda$, say by a
factor of three, can increase or decrease the cross-section by one order of
magnitude.}  
Bearing the vanishingly small cross-section for this reaction 
in the SM (due to GIM suppression), 
the large cross-section predicted in Model~III is 
especially gratifying. Therefore, this reaction may serve as 
a unique test of the SM and, in particular, of its GIM mechanism.
From the experimental point of view, it should be emphasized that although
$\sigma^{eetc}$ is found to be one order of magnitude smaller 
then $\sigma^{\nu
\nu tc}$, the $t \bar c e^+e^-$ signature may be 
easier to detect as it does not have the missing energy associated
with the two neutrinos in the $t \bar c \nu_e {\bar {\nu}}_e$ final state.
 Moreover, at $\sqrt s \gsim 1$ TeV, the 
$t \bar c \nu_e {\bar {\nu}}_e$ and $t \bar
c e^+ e^-$ signatures are to some extent unique, as 
other simple FC $s$-channel
processes like $e^+e^- \to Z \to t \bar c$, $e^+e^- \to Z{\cal H} \to Z t \bar
c$ and $e^+e^- \to A {\cal H} \to t \bar t c \bar c, t \bar c f \bar f$ 
tend to drop as $1/s$ and are therefore expected to yield much smaller
production rates at an $e^+e^-$ collider with $\sqrt{s} \gsim1$ TeV\null.

It should be also noted that we do not anticipate serious background 
problems for this reaction, since it will be difficult 
to fake a $t \bar c$ event with other ``normal'' modes, 
like the $t \bar t$ and the $W^+W^-$ final states, which 
may have higher production rates 
then the $t \bar c \nu_e {\bar\nu}_e$ mode \cite{prl79p1217}.     
For example, we have shown in \cite{prl79p1217}, 
that, in Model III, the $t \bar c \nu_e {\bar\nu}_e$ final state may 
come out 
favorable compared to the normal $t \bar t \nu_e {\bar\nu}_e$, 
in the range where $\sigma^{\nu \nu t c} \gsim 1$ fb.

Finally, in Model III, the same tree-level 
$VV \to t \bar c$ amplitude may give rise to enhanced 
decay and production rates of the rare three-body top decays 
$t \to W^+ W^- c,ZZc$ and the $Zt \bar c$ final state 
through $e^+ e^- \to Z t \bar c$ 
(for more detail see \cite{prl79p1217}). \\    

\noindent {4. \bf General Concluding Remarks} \\

To conclude this talk let us add two additional remarks:

\begin{itemize}

\item It is most likely that the Higgs particles, if at all present,
will have been discovered by the time the NLC starts its first
run. If indeed such a particle is detected 
with a mass of a few hundreds GeV,
it will be extremely important to investigate the
reactions $e^+ e^- \to t \bar t h$, $e^+ e^- \to t \bar t Z$, 
$e^+e^- \to t \bar c \nu_e {\bar {\nu}}_e$ and
$e^+e^- \to t \bar c e^+ e^-$ in the NLC, as they may serve as strong
evidence for the existence of a nonminimal scalar sector with CP-violating 
and/or FC
scalar couplings to fermions. Needless to say, such signatures 
of CP-violation and FCS interactions, residing in the scalar potential,
 will 
rule out the Minimal Supersymmetric Standard Model (MSSM).  
In addition, since supersymmetry strongly
disfavors an $h$ heavier than $ \sim 150$ GeV, the
detection of a Higgs particle above this 
limit would drive the study of general
extended scalar sector, not of a supersymmetric origin,
and in turn, this should encourage the study of CP-nonconserving
 and FC
effects such as the ones presented in this talk.

\item It should be emphasized that the large 
CP-violating effects found in 
$e^+ e^- \to t \bar t h$ and $e^+ e^- \to t \bar t Z$  
for Model II can be generalized  to model III as well.
In Model~III, the CP-odd phase in the ${\cal H} t \bar t$ vertex can arise
from a phase in the Yukawa couplings  $U^2_{ij}$ and $D^2_{ij}$ defined in
Eq.~\ref{yukawa}. With an appropriate choice 
of parameters, a comparable tree-level CP-odd effect 
may arise within Model~III in 
$e^+ e^- \to t \bar t h$ 
and $e^+ e^- \to t \bar t Z$ (see also \cite{hepph9707284}). 
Thus, detection
or no detection of the FC channels such as 
$e^+e^- \to t \bar c \nu_e {\bar {\nu}}_e$ and
$e^+e^- \to t \bar c e^+ e^-$ discussed here, 
along with evidence for CP-violation
in the Higgs sector that could emanate in 
the reactions $e^+ e^- \to t \bar t h$ and $e^+ e^- \to t \bar t Z$
in high energy $e^+e^-$ colliders, may well be the only
way  to experimentally distinguish between scalar dynamics of a Model~II 
or a Model~III origin.
\end{itemize}

\end{document}